\documentclass[a4paper,10pt,twoside]{cpc-hepnp}

\usepackage{multicol}
\usepackage{graphicx}
\usepackage{booktabs}
\usepackage{amssymb,bm,mathrsfs,bbm,amscd}
\usepackage[tbtags]{amsmath}
\usepackage{lastpage}

\begin{document}

\fancyhead[c]{\small Submitted to ``Chinese Physics C''} \fancyfoot[C]{\small 010201-\thepage}

\footnotetext[0]{Received 31 Jul 2015}

\title{Beam size and position measurement based on logarithm processing algorithm in HLS II\thanks{Supported by National Natural Science
Foundation of China (11005105, 11175173) }}

\author{%
      CHENG Chao-Cai(³Ì³¬²Å)$^{1)}$\email{ccc1989@mail.ustc.ac.cn}%
\quad SUN Bao-Gen(ËïÝá¸ù)$^{2)}$\email{bgsun@ustc.edu.cn}%
\quad YANG Yong-Liang(ÑîÓÀÁ¼)
\quad ZHOU Ze-Ran(ÖÜÔóÈ»)\\
\quad LU Ping(¬ƽ)
\quad WU Fang-Fang(Îâ·¼·¼)
\quad WANG Ji-Gang(Íõ¼¾¸Õ)
\quad TANG Kai(ÌÆ¿­)\\
\quad LUO Qing(ÂÞóä)
\quad LI Hao(Àîð©)
\quad ZHENG Jia-Jun(Ö£¼Ñ¿¡)
\quad DUAN Qing-Ming(¶ÎÇàÃ÷)
}
\maketitle

\address{%
National Synchrotron Radiation Laboratory, University of Science and Technology of China, Hefei, 230029, China
}

\begin{abstract}
A logarithm processing algorithm to measure beam transverse size and position
is proposed and preliminary experimental results in Hefei Light Source II (HLS II)
are given. The algorithm is based on only 4 successive channels of 16 anode channels
of multianode photomultiplier tube (MAPMT) R5900U-00-L16 which has typical rise time
of 0.6 ns and effective area of 0.8$\times$16 mm for a single anode channel. In the paper,
we firstly elaborate the simulation results of the algorithm with and without channel
inconsistency. Then we calibrate the channel inconsistency and verify the
algorithm using general current signal processor Libera Photon in low-speed
scheme. Finally we get turn-by-turn beam size and position and calculate the
vertical tune in high-speed scheme. The experimental results show that measured
values fit well with simulation results after channel differences are calibrated and
the fractional part of the tune in vertical direction is 0.3628 which is very
close to the nominal value 0.3621.
\end{abstract}

\begin{keyword}
logarithm processing algorithm, beam size and position, multianode photomultiplier tube, channel inconsistency
\end{keyword}

\begin{pacs}
29.20.db, 29.85.Ca, 29.90.+r
\end{pacs}

\footnotetext[0]{\hspace*{-3mm}\raisebox{0.3ex}{$\scriptstyle\copyright$}2013
Chinese Physical Society and the Institute of High Energy Physics
of the Chinese Academy of Sciences and the Institute
of Modern Physics of the Chinese Academy of Sciences and IOP Publishing Ltd}%

\begin{multicols}{2}

\section{Introduction}
In electron storage ring and collider complex, beam transverse profile measurement
based on visible light from synchrotron radiation (SR) is widely used \cite{lab1}. In general,
CCD can be applied to measure beam size and position steadily and online. But
conventional CCD has a long exposure time and very low frame rate. In HLS II,
there has been installed a CCD camera to measure beam profile and monitor the
machine status. The CCD camera has the maximum frame rate of 205 Hz at the full
resolution (640$\times$480), and the electron beam go through about 22k turns
within single frame interval. As a result, the measurement results are indeed
the average of multiple bunches in multiple turns.

In order to track the beam size and position of single bunch, it is necessary
to acquire transient light intensity information utilizing high speed photodetector
and data acquisition system to ensure that the transformed light pulse are
not piled up along the pulse train. The typical schemes are VEPP-4M \cite{lab2} and CESR-c \cite{lab3}
which calculate the turn-by-turn or bunch-by-bunch beam size and position to study beam-beam
interaction, beam dynamics, etc. In these measurement schemes, dozens of channel information
up to 16 or 32 are sampled in parallel to fit the shape of Gaussian beam spot,
which increase the costs and complexity. Therefore, we put forward a logarithm
processing algorithm to calculate the beam size and position which needs only 4
successive channels, and carry out some experiments to verify the algorithm.

\section{Principle of the measurement system}
At the moment there are 4 applications in the visible diagnostic beam line
at $15^{\circ}$ outlet of bending magnet in HLS II. Fig.~\ref{fig1} shows the block diagram of
the measurement system for CCD and vertical MAPMT. Synchrotron radiation is first
reflected by water-cooled Oxygen Free Copper (OFC) mirror in vacuum chamber and
then the visible light pass through several mirrors and a main focusing lens
with magnification of 1 before reaching the optical table. MAPMT and CCD \cite{lab4} share
the same magnification of 2 by front-end optical path. Then the beam spot reaches
the horizontal and vertical MAPMT cathode with magnification of 6 and magnification
of 24 respectively. Magnification can be altered to meet the need in different
machine status. The two MAPMTs are fixed on three-dimensional translation stage
so as to find the image plane conveniently. Front-end circular slit constraints
the opening angle of SR within $\pm$3.8 mrad to decrease the depth of field error and diffraction
error \cite{lab5}. A band pass filter with 500 nm peak wavelength and 10 nm full width at half
maximum (FWHM) is used to allow quasi-monochromatic light to get to the sensitive area of MAPMT.
\end{multicols}
\begin{center}
\includegraphics[width=12cm]{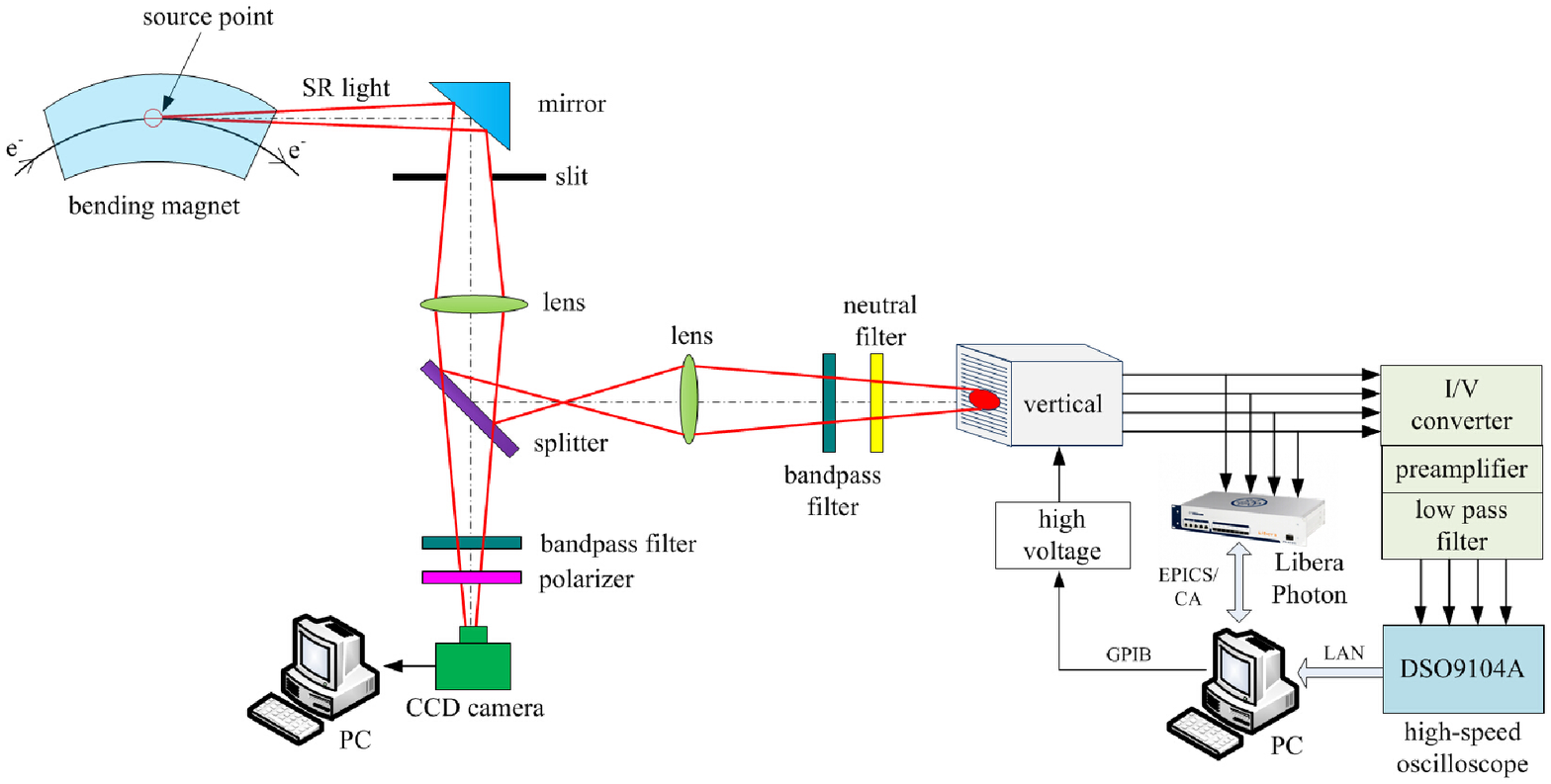}
\figcaption{\label{fig1} Block diagram of the measurement system.}
\end{center}
\begin{multicols}{2}
The measurement system works in two schemes. In low-speed scheme, photocurrent of 4
successive channels is sampled by general current signal processor Libera Photon
from Instrumentation Technologies and then sent to remote PC by EPICS/CA protocol,
where we can calibrate the channel inconsistency, implement the logarithm processing
algorithm to extract beam size and position, and control the negative high voltage
to protect the MAPMT. In high-speed scheme, photocurrent is firstly conditioned by
analog front-end electronics (AFE) including current to voltage conversion, preamplifier
and low pass filter. Then transformed signals are sampled simultaneously by high speed
oscilloscope DSO9104A and acquired over Local Area Network (LAN) by remote
PC where we can calculate turn-by-turn beam size and position,
perform Fast Fourier Transformation (FFT), etc.

\section{Logarithm processing algorithm \cite{lab6}}

As SR light irradiated by electron beam from bending magnet in
storage ring presents Gaussian distribution, we assume SR light intensity
distribution function $\Phi(x)$ as Eq.~(\ref{eq1}).
\begin{eqnarray}
\label{eq1}
\Phi(x)=\Phi_0\exp\left[ -{\left(\frac{x-\delta}{\sqrt{2}\sigma}\right)}^2\right].
\end{eqnarray}
where $\Phi_0$ is the maximum light intensity; $\delta$ is beam displacement that center
of the beam spot offsets specified center of the MAPMT; $\sigma$ is beam size.

Based on the assumption of Gaussian distribution, the SR light intensity
distribution at MAPMT is shown in Fig.~\ref{fig2}. The rectangular region of
$I_1$, $I_2$, $I_3$ and $I_4$ represents effective photosensitive areas with
0.2 mm gap between them.
\begin{center}
\includegraphics[width=6cm]{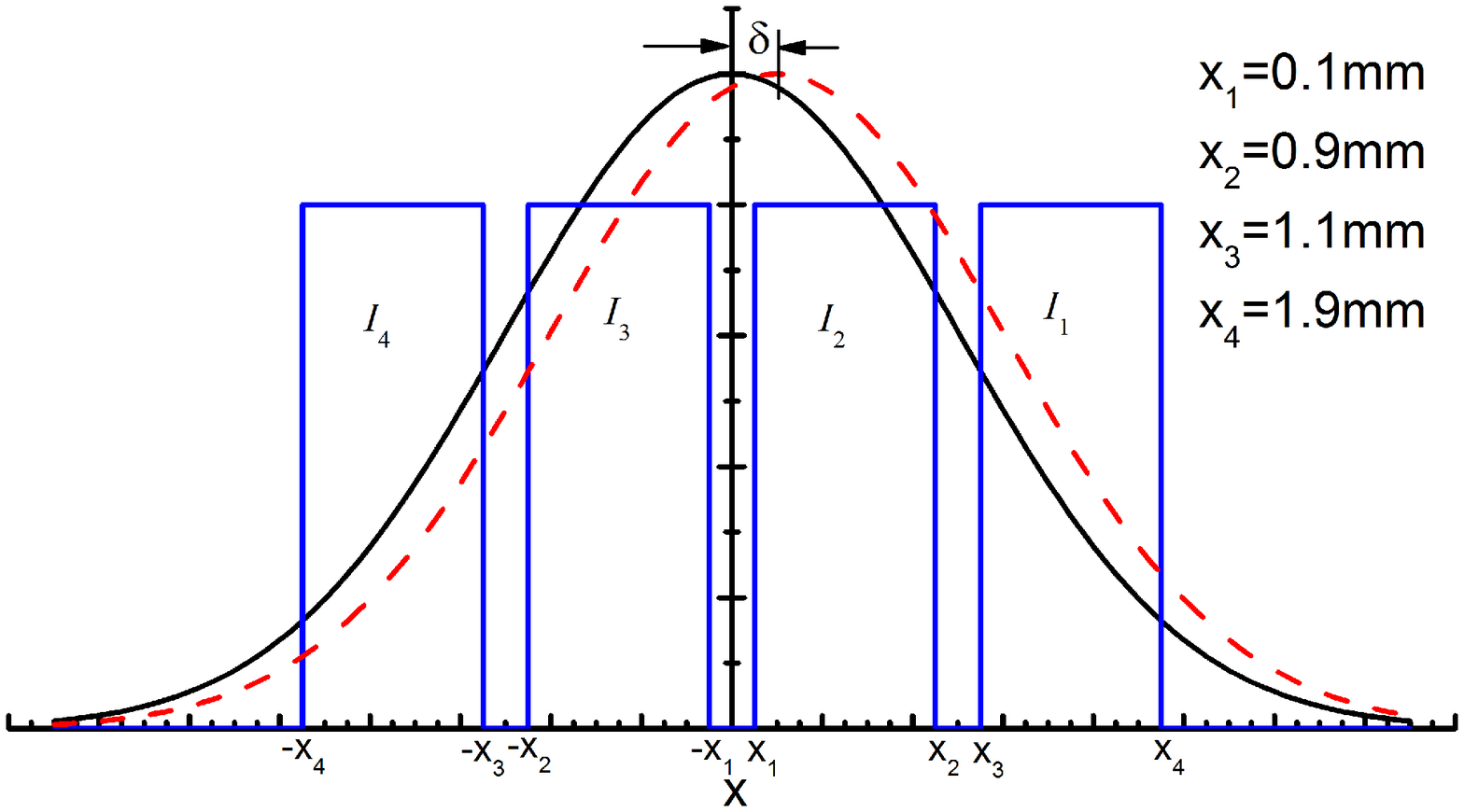}
\figcaption{\label{fig2} Light intensity distribution at MAPMT.}
\end{center}

Photocurrent of $I_1$, $I_2$, $I_3$ and $I_4$, which derives from 4 successive channels,
is proportional to the integral of light intensity and can be expressed by the subtraction
of two error functions shown in Eq.~(\ref{eq2}).
\begin{eqnarray}
\label{eq2}
\begin{cases}
\displaystyle I_1=\vint\nolimits_{x_3}^{x_4} \Phi(x){\rm{d}} x=I_0\left[{\rm erf}(\frac{x_4-\delta}{\sqrt{2}})-{\rm erf}(\frac{x_3-\delta}{\sqrt{2}})\right],\\
\displaystyle I_2=\vint\nolimits_{x_1}^{x_2} \Phi(x){\rm{d}} x=I_0\left[{\rm erf}(\frac{x_2-\delta}{\sqrt{2}})-{\rm erf}(\frac{x_1-\delta}{\sqrt{2}})\right],\\
\displaystyle I_3=\vint\nolimits_{-x_2}^{-x_1} \Phi(x){\rm{d}} x=I_0\left[{\rm erf}(\frac{-x_1-\delta}{\sqrt{2}})-{\rm erf}(\frac{-x_2-\delta}{\sqrt{2}})\right],\\
\displaystyle I_4=\vint\nolimits_{-x_4}^{-x_3} \Phi(x){\rm{d}} x=I_0\left[{\rm erf}(\frac{-x_3-\delta}{\sqrt{2}})-{\rm erf}(\frac{-x_4-\delta}{\sqrt{2}})\right].
\end{cases}
\end{eqnarray}
where ${\rm erf}(x)=\frac{2}{\sqrt{\pi}}\int_0^x{\rm e}^{-t^2}{\rm d}t$; $x_1$=0.1 mm, $x_2$=0.9 mm, $x_3$=1.1 mm and $x_4$=1.9 mm;
$I_0$ is the maximum anode photocurrent.

\subsection{Ideal case}

In ideal condition when the 4 continuous channels have the same response characteristic,
we define ideal size signal $S_{\ln}(\sigma,\delta)$ and ideal position signal $P_{\ln}(\sigma,\delta)$ in Eq.~(\ref{eq3}).
\begin{eqnarray}
\label{eq3}
\begin{cases}
\displaystyle S_{\ln}(\sigma,\delta) =  {\left[\ln (I_2I_3/I_1I_4)\right]}^{ -1/2},\\
\displaystyle P_{\ln}(\sigma,\delta) =  \frac{\ln (I_1I_2/I_3I_4)}{\ln (I_2I_3/I_1I_4)}.
\end{cases}
\end{eqnarray}

Substituting the integral result shown in Eq.~(\ref{eq2}) into Eq.~(\ref{eq3}) and simplifying it, we can
obtain how the size and position signal vary from size and position illustrated in Fig.~\ref{fig3}.
When beam size increases from 0.2 mm to 2 mm with 0 mm beam displacement, size signal
$S_{\ln}(\sigma,\delta)$ has a good linear relation with size $\sigma$ and the linear fitting equation can be
expressed by $S_{\ln}(\sigma)=0.03924+0.69\sigma$. When beam size varies within $0.8 \rm{mm}\sim\rm{2 mm}$, beam
displacement $\delta$ has an effect on normalized size signal within 1\%. Meanwhile, position
signal $P_{\ln}(\sigma,\delta)$ has a good linear relation with beam displacement when beam displacement
varies from --2 mm to 2 mm and linear fitting equation is $P_{\ln}(\delta)=2\delta$. Furthermore, beam
size has an impact on position signal sensitivity within 1\%.
\end{multicols}
\begin{center}
\includegraphics[width=12cm]{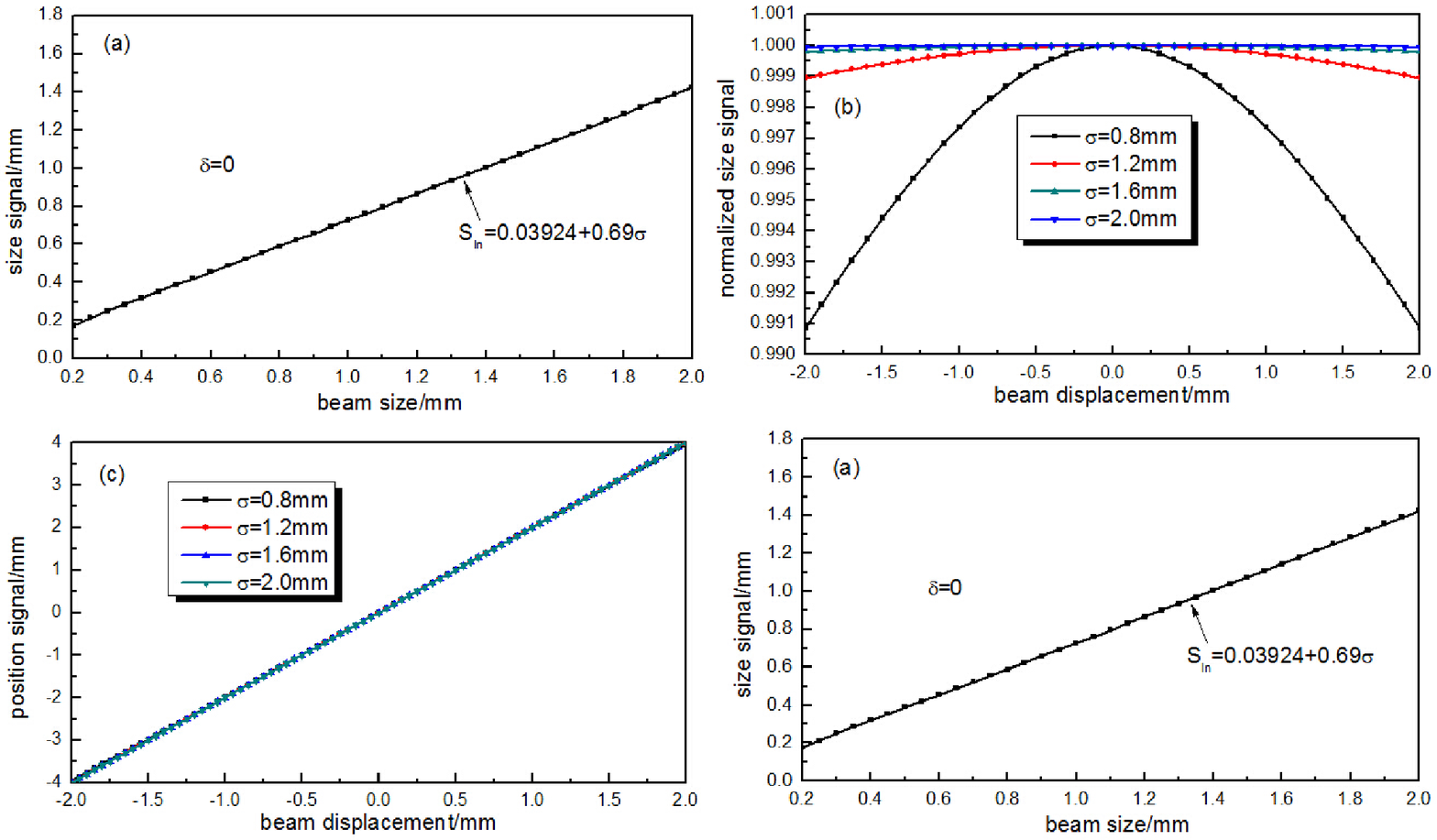}
\figcaption{\label{fig3} The simulation results of ideal beam size and position signal in different conditions:
(a) size signal versus beam size when $\delta$=0 mm;
(b) normalized size signal versus beam displacement when $\sigma=0.8\sim2.0 \rm{mm}$;
(c) position signal versus beam displacement when $\sigma=0.8\sim2.0 \rm{mm}$;
(d) position signal sensitivity versus beam size.}
\end{center}
\begin{multicols}{2}
From the above simulation results, we can get the approximate equations expressed in Eq.~(\ref{eq4}).
\begin{eqnarray}
\label{eq4}
\begin{cases}
\displaystyle S_{\ln}(\sigma,\delta)\approx S_{\ln}(\sigma)=0.03924+0.69\sigma,\\
\displaystyle P_{\ln}(\sigma,\delta)\approx P_{\ln}(\sigma)=2\delta.
\end{cases}
\end{eqnarray}

\subsection{Non-ideal case}

Given the channel inconsistency resulting from anode uniformity and cross-talk among channels,
we introduce the channel gain factor $g_i(i=1,\;2,\;3,\;4)$ to correct channel inconsistency based on
linear approximation, as is shown in Eq.~(\ref{eq5}).
\begin{eqnarray}
\label{eq5}
\begin{cases}
\displaystyle \hat{S}_{\ln}(\sigma,\delta)={\left[\ln (\hat{I}_2 \hat{I}_3 / \hat{I}_1 \hat{I}_4)\right]}^{-1/2}\\
\hspace{37pt}={\left[ \ln (g_2g_3/g_1g_4)+\ln (I_2I_3/I_1I_4)\right]}^{-1/2},\\
\displaystyle \hat{P}_{\ln}(\sigma,\delta)=\frac{\ln (\hat{I}_1 \hat{I}_2 / \hat{I}_3 \hat{I}_4)}{\ln (\hat{I}_2 \hat{I}_3 / \hat{I}_1 \hat{I}_4)}\\
\hspace{37pt}=\displaystyle {\frac{\ln (g_1g_2/g_3g_4)+\ln (I_1I_2/I_3I_4)}{\ln (g_2g_3/g_1g_4)+\ln (I_2I_3/I_1I_4)}}.
\end{cases}
\end{eqnarray}
where $\hat{S}_{\ln}(\sigma,\delta)$ is actual size signal, $\hat{P}_{\ln}(\sigma,\delta)$ is actual position signal,
$\hat{I}_i=g_iI_i(i=1,\;2,\;3,\;4)$ represents actual anode photocurrent from 4 successive channels with channel inconsistency.

If we define $\displaystyle d_s=\ln (g_2g_3/g_1g_4)$, $\displaystyle d_p=\ln (g_1g_2/g_3g_4)$ and put
Eq.~(\ref{eq4}) into Eq.~(\ref{eq5}), we can get:
\begin{eqnarray}
\label{eq6}
\begin{cases}
\ \hat{S}_{\ln}(\sigma,\delta)\approx S_{\ln}(\sigma){\left[1+\tilde{d}_sS^2_{\ln}(\sigma)\right]}^{-1/2},\\
\ \hat{P}_{\ln}(\sigma,\delta)\approx P_{\ln}(\delta){\left[1+\tilde{d}_sS^2_{\ln}(\sigma)\right]}^{-1}+\\
\hspace{50pt} \tilde{d}_pS^2_{\ln}(\sigma){\left[1+\tilde{d}_sS^2_{\ln}(\sigma)\right]}^{-1}.
\end{cases}
\end{eqnarray}
where $\tilde{d}_s$ and $\tilde{d}_p$ are gain correction factors.

According to Eq.~(\ref{eq6}), we know that actual position signal has a good linear relation with
beam position, but actual size signal presents a huge nonlinearity with size. In addition,
$\tilde{d}_s$ and $\tilde{d}_p$ can be induced by slope and intercept of $\hat{P}_{\ln}(\sigma,\delta)$ if beam size and magnification
can be acquired in advance. Hence, modified size signal $\tilde{S}_{\ln}(\sigma,\delta)$  and modified
position signal $\tilde{P}_{\ln}(\sigma,\delta)$ can be expressed in Eq.~(\ref{eq7}).
\begin{eqnarray}
\label{eq7}
\begin{cases}
\displaystyle \tilde{S}_{\ln}(\sigma,\delta)= {\left[ \ln (\hat{I}_2\hat{I}_3/\hat{I}_1\hat{I}_4)-\tilde{d}_s\right]}^{-1/2},\\
\displaystyle \tilde{P}_{\ln}(\sigma,\delta)= \frac{\ln (\hat{I}_1\hat{I}_2/\hat{I}_3\hat{I}_4)-\tilde{d}_p}
{\ln (\hat{I}_2\hat{I}_3/\hat{I}_1\hat{I}_4)-\tilde{d}_s}.
\end{cases}
\end{eqnarray}

\section{Experimental results}
\subsection{Low-speed scheme}

In low-speed scheme, the horizontal measurement results are described. In order to simulate
the movement of beam center, we move the three-dimensional translation stage for horizontal
MAPMT horizontally. The movement range is from 13.0 mm to 16.8 mm with 0.1 mm step.
The process of movement is very short, so the beam size measured by CCD remains unchanged.

Figure ~\ref{fig4} shows how the average anode output current varies with the moved position.
CH4, CH5, CH6 and CH7 are corresponding to channel number 4, 5, 6 and 7 respectively.
When the horizontal translation stage increases, average anode current of CH4 and CH5
increases and it is just opposite for CH6 and CH7, which roughly indicates the beam spot
center is moving from CH7 to CH4. In the middle point of 14.9 mm, anode current of CH4
and CH7 are approximate the same, and so does it for CH5 and CH6. Thus this position can
be thought to be the absolute center.
\begin{center}
\includegraphics[width=6cm]{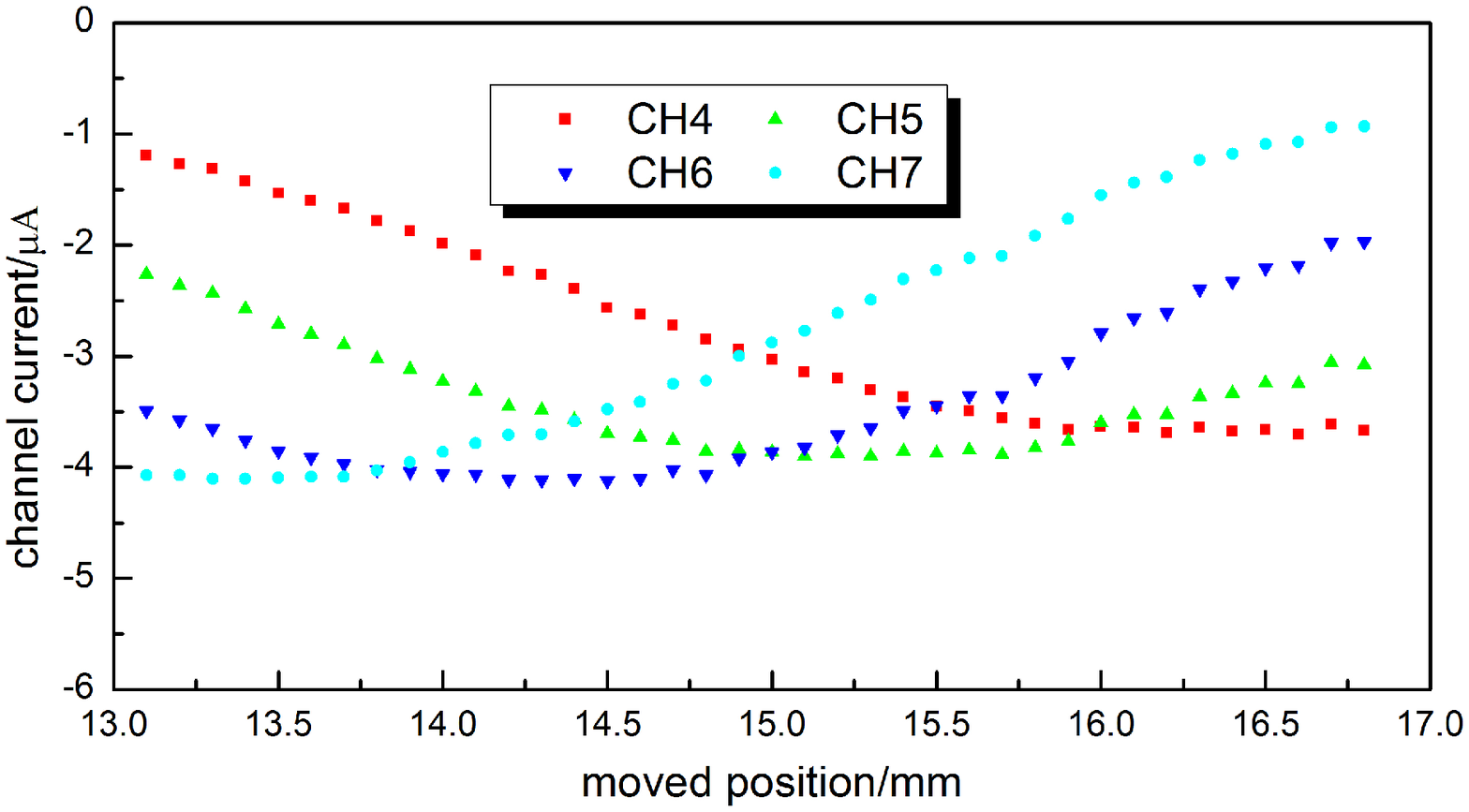}
\figcaption{\label{fig4} Average anode current of 4 successive channels.}
\end{center}

Putting average anode current into Eq.~(\ref{eq5}), we can get how the actual position signal varies with the
simulated beam displacement shown in Fig.~\ref{fig5}. The horizontal coordinate axis has
been adjusted by subtracting the absolute center. Applying linear fitting to
the raw discrete points, we can get the fitting equation with slope of 1.7894
and intercept of --0.1206. Actual position signal has a linear relation with
simulated beam displacement as described in Eq.~(\ref{eq6}). However, the fitting curve
does not cross the point (0, 0), and slope does not equal to the simulation
result in ideal case shown in Fig.~\ref{fig3}(c). The difference mainly comes from
the channel inconsistency and must be calibrated.
\begin{center}
\includegraphics[width=6cm]{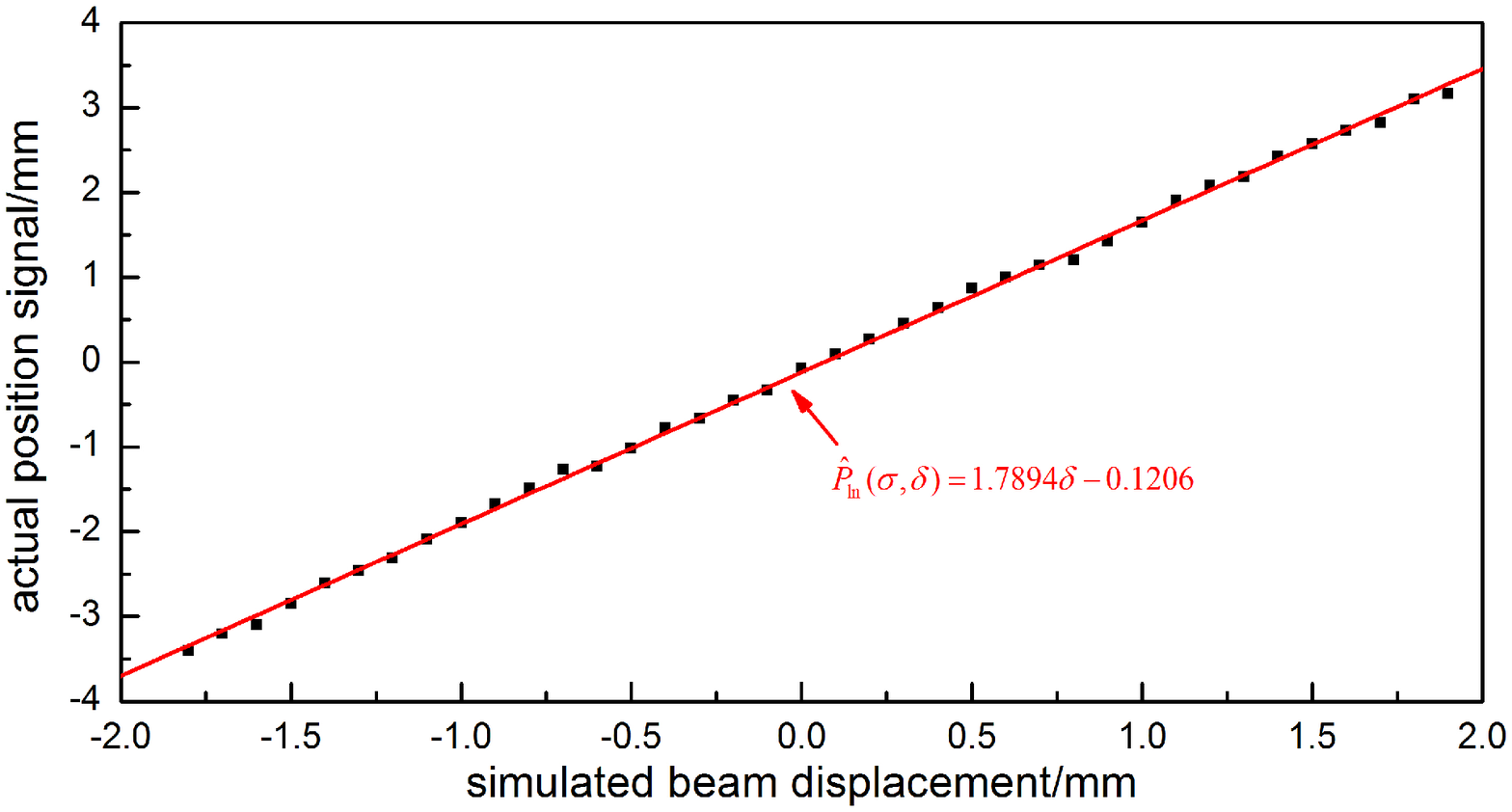}
\figcaption{\label{fig5} Actual position signal with channel inconsistency.}
\end{center}

Substituting the magnifying beam size imaged at the horizontal MAPMT cathode
into Eq.~(\ref{eq6}), we can get the measured gain correction factors
$\tilde{d}_s$=0.0590 and $\tilde{d}_p$=--0.0675 on the basis of the fitting equation
shown in Fig.~\ref{fig5}. As a consequence, the relationship between the modified position
and simulated position can be illustrated in Fig.~\ref{fig6}. Linear fitting results show that
the modified position signal has a good linear relation with simulated beam displacement
and the fitting curve cross point (0, 0) with slope of 2.002 and intercept of --0.008 which
agrees well with the simulation result in Fig .3(c).
\begin{center}
\includegraphics[width=6cm]{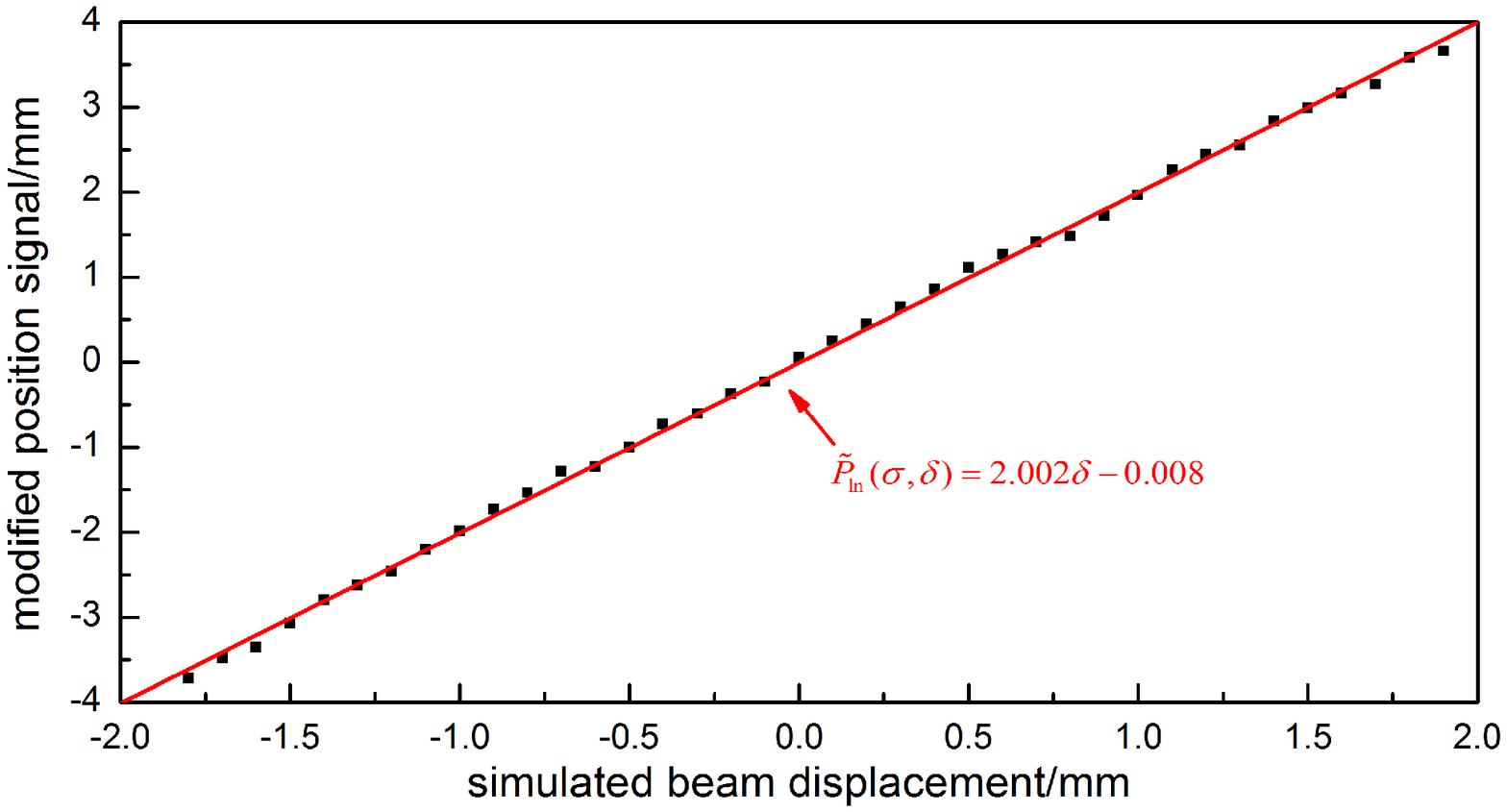}
\figcaption{\label{fig6} Modified position signal.}
\end{center}

Beam size calculated by direct equation in Eq.~(\ref{eq5}) and modified equation in Eq.~(\ref{eq7}) is
illustrated in Fig.~\ref{fig7} respectively. Direct and modified beam size gradually decreases
when beam center is moved from --1.8 mm to 1.9 mm although the modified beam size signal
which is proportional to beam size should remain almost unchanged according to Fig .3(b). But the
modified beam size is more close to the nominal value 0.332 mm measured by CCD and is more reliable.
\begin{center}
\includegraphics[width=6cm]{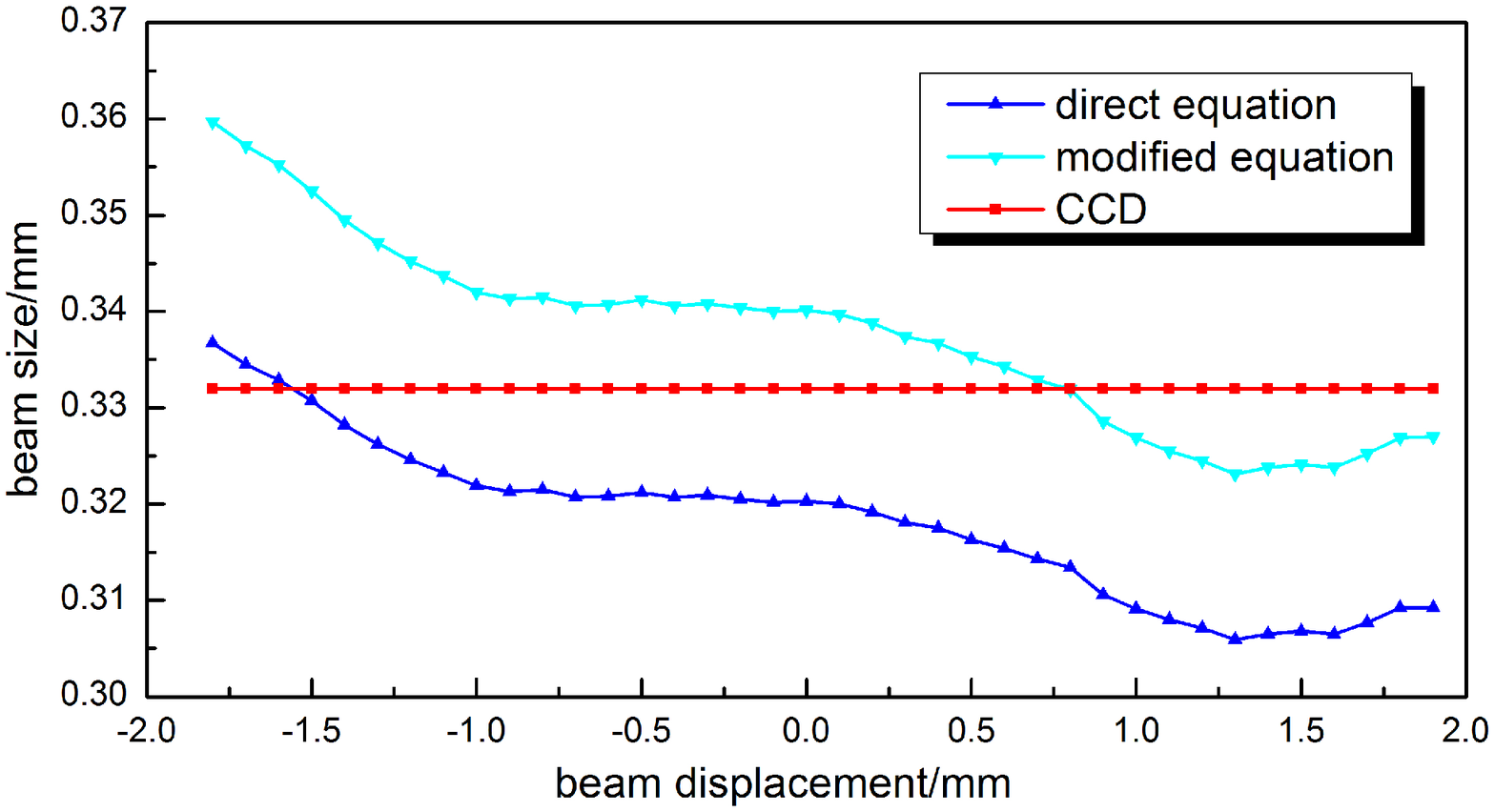}
\figcaption{\label{fig7} Direct and modified beam size.}
\end{center}

\subsection{High-speed scheme}
In high-speed scheme, we take the vertical measurement results for example.
Single bunch fills in the storage ring with 3 mA beam current and electron beam is
excited by swept frequency signal which comes from the tracking generator module in the spectrum analyzer.
Signals from AFE are continuously sampled 500,000 points with 10 GS/s sample rate for 4 channels.
External trigger signal with frequency of 4.534 MHz synchronized with the revolution frequency
is applied to keep pace with the pulse signal produced by SR light. Fig.~\ref{fig8} shows the
peak values of each turn extracted from raw data points with peak search algorithm.
As the sample points per one period is about 2200, we can sample 227 turns data in total.
CH8, CH9, CH10 and CH11 are corresponding to channel number 8, 9, 10 and 11 respectively.
Peak value of CH9 and CH10 is larger than CH8 and CH11 indicating that beam center is
located between the CH9 and CH10 which satisfies the conditions of the algorithm.
\begin{center}
\includegraphics[width=6cm]{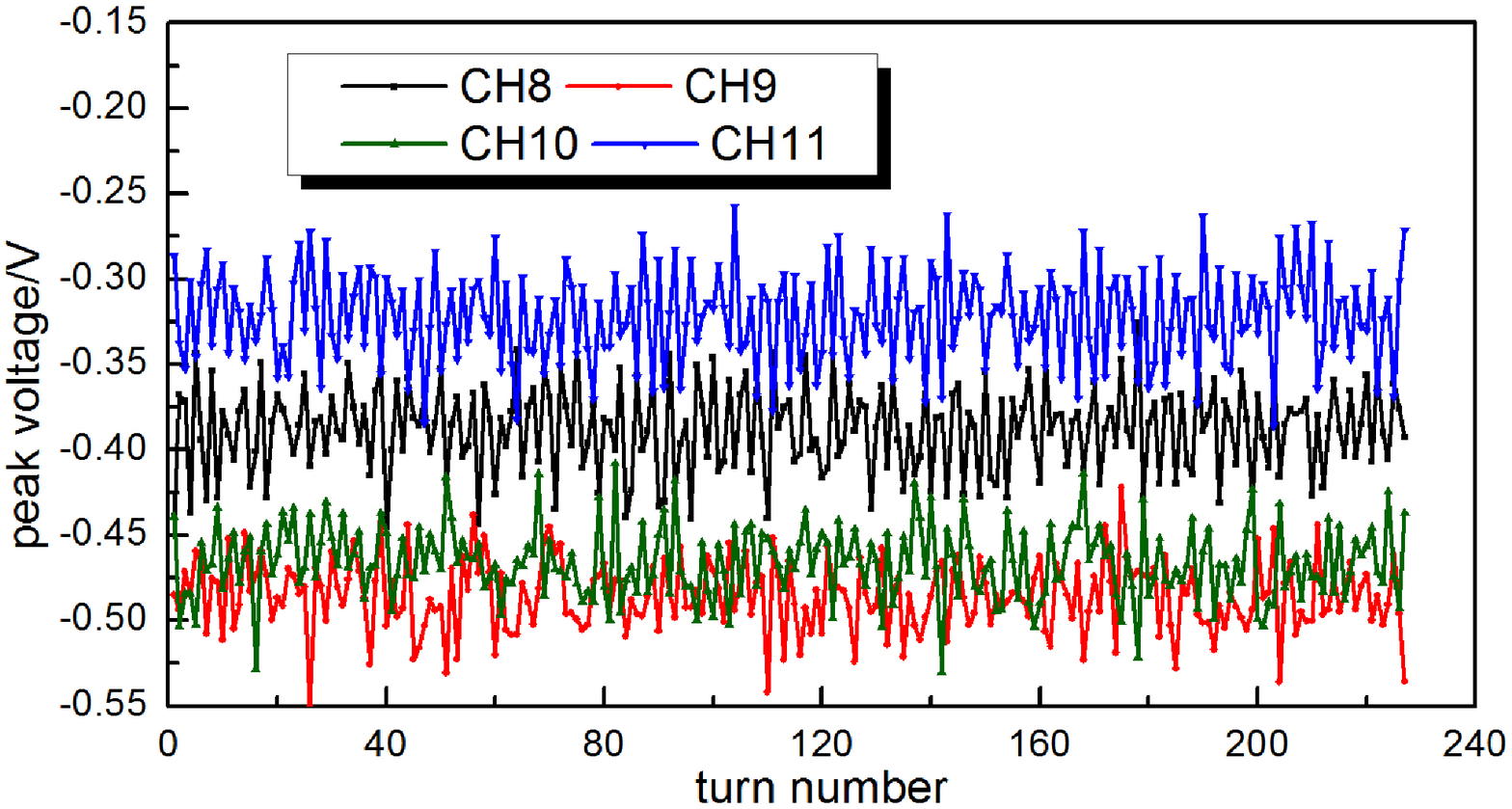}
\figcaption{\label{fig8} Peak values of each turn for 4 successive channels.}
\end{center}

Applying FFT to all the 500,000 points raw data and enlarging the view of spectrum of CH9
in frequency range of 0 Hz to 10 MHz shown in Fig.~\ref{fig9}, we can clearly
see vertical synchronous oscillation sidebands appear at frequency of 1.620 MHz,
2.900 MHz, 6.160 MHz and 7.440 MHz except for two main harmonic components
at the frequency of 4.530 MHz and 9.060 MHz which are roughly equal to the
integral multiple of revolution frequency. These sideband elements result
from the interactions between the synchronous light and slit. In HLS II,
frequency at 1.620 MHz and 6.160 MHz correspond to betatron oscillation
in vertical direction, and frequency at 2.900 MHz and 7.440 MHz are the
mirrored components of vertical betatron oscillation.
\begin{center}
\includegraphics[width=6cm]{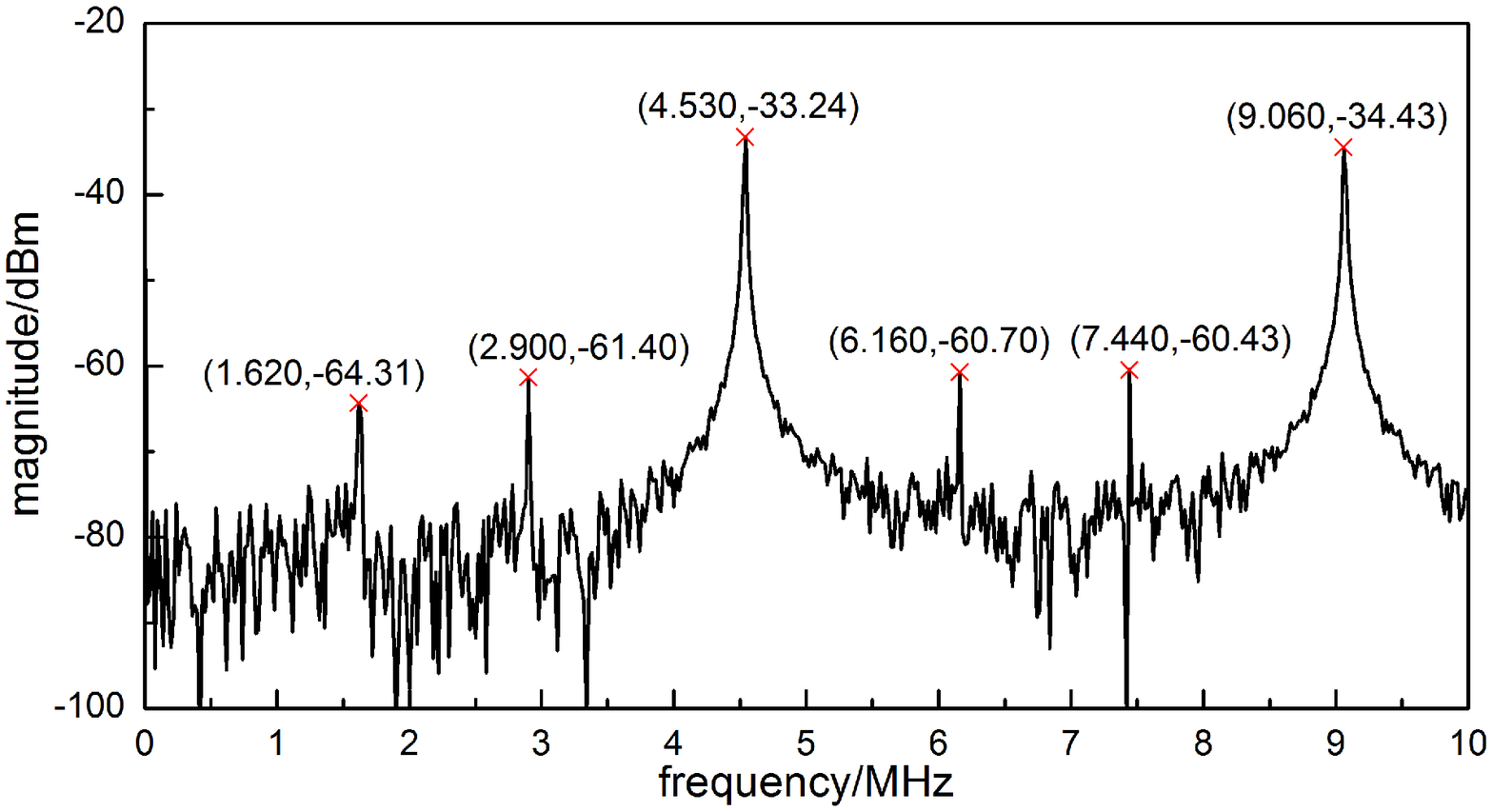}
\figcaption{\label{fig9} Spectrum of CH9.}
\end{center}

With the channel inconsistency being calibrated and logarithm
processing algorithm being adopted with 227 turns peak values, we can get the
turn-by-turn beam size and position shown in Fig.~\ref{fig10}.
From Fig.~\ref{fig10}(a), beam size varies between 64.5 $\mu$m and 93.5 $\mu$m with the
average of 76.4 $\mu$m. From Fig.~\ref{fig10}(b), beam position varies between --8.2 $\mu$m
and 22.6 $\mu$m and the average value is 8.0 $\mu$m . Meanwhile,
we record the beam size with CCD which tells that the
beam size is about 70 $\mu$m. Our average result of beam size
is larger than the beam size from CCD for about 8\% which
may result from the amplification of light path.
\begin{center}
\includegraphics[width=6cm]{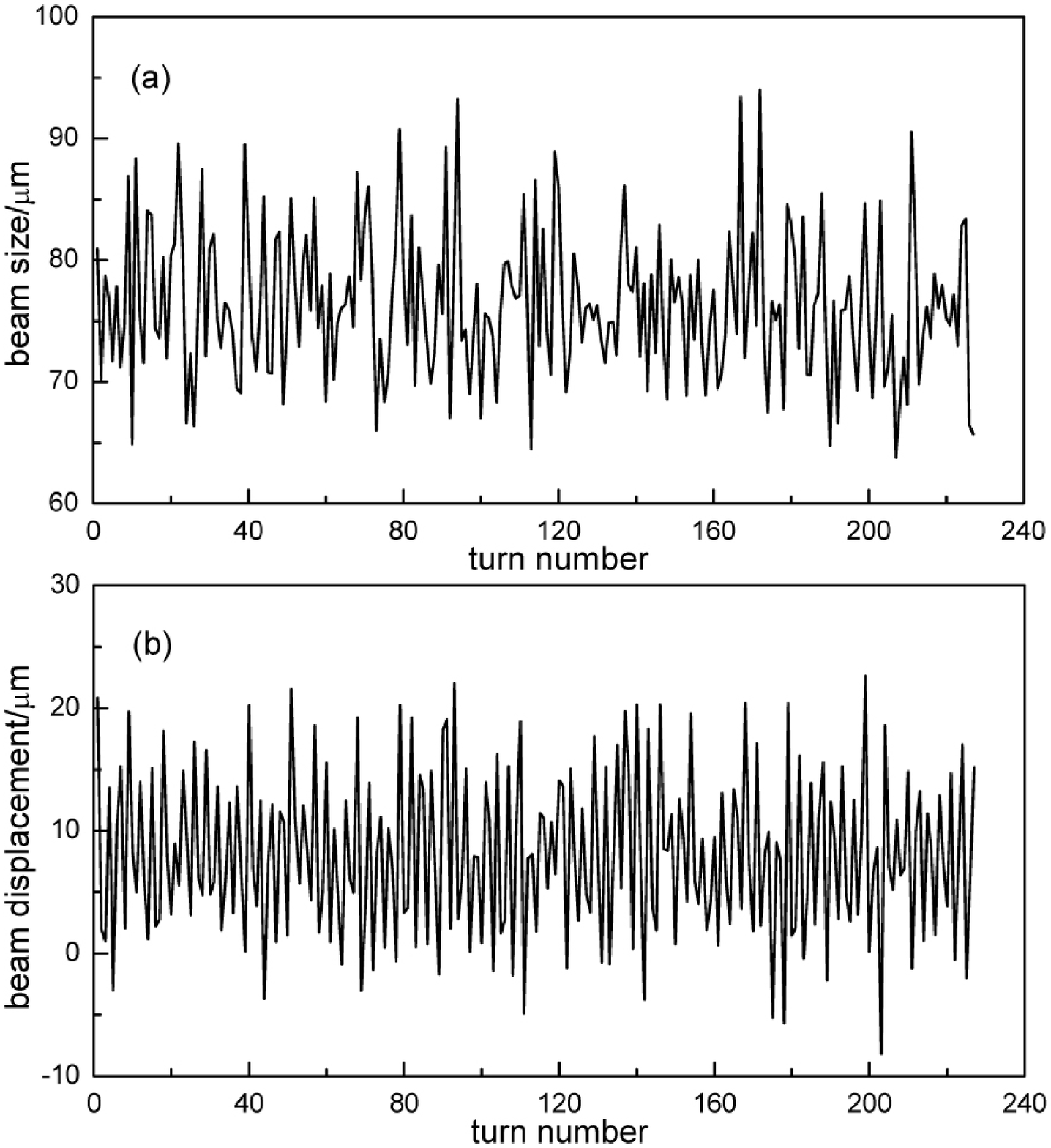}
\figcaption{\label{fig10} Beam size and position for 227 turns: (a) beam size; (b) beam position.}
\end{center}

Applying FFT to the turn-by-turn beam position and getting rid
of the dc component, we can obviously see a peak signal at the
frequency of 1.645 MHz corresponding to the vertical betatron
oscillation sideband shown in Fig.~\ref{fig11}. So the vertical fractional
tune is 0.3628 with 4.534 MHz revolution frequency. The measured
result is in good agreement with the nominal value 0.3621.
\begin{center}
\includegraphics[width=6cm]{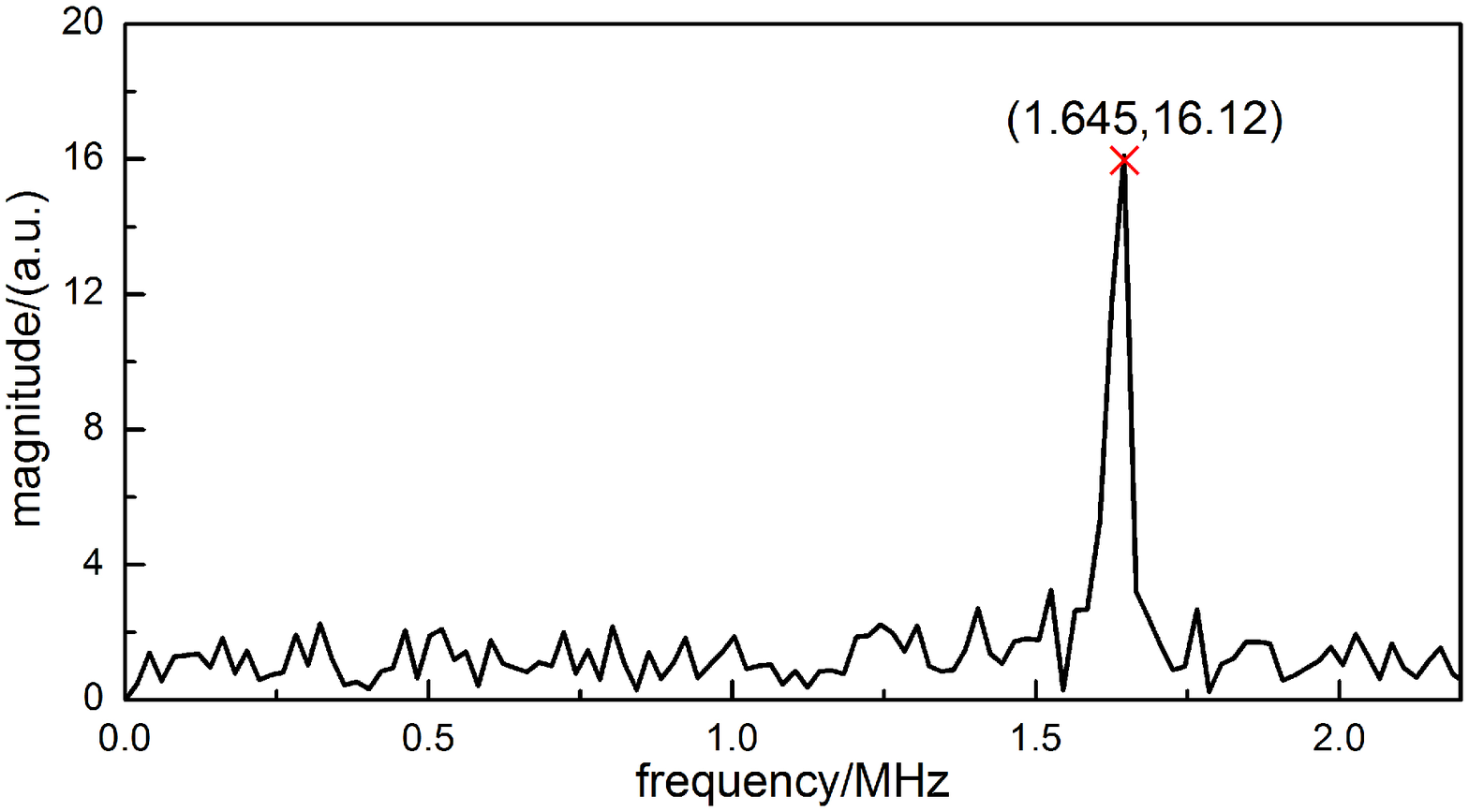}
\figcaption{\label{fig11} Spectrum of turn-by-turn position.}
\end{center}

\section{Conclusion}

A logarithm processing algorithm used to calculate beam transverse
size and position based on SR light in HLS II is developed and preliminary
experimental results are given. The measurement
results are in good agreement with the simulation results after the
channel inconsistency is calibrated. Vertical fractional tune of 0.3628
is obtained according to the turn-by-turn beam position which is very
close to the nominal value 0.3621. More experiments will be conducted
in multiple bunches filling mode, and effective ways to excite the
obvious change of beam size should be applied to study beam
instability and beam dynamics.

\end{multicols}

\clearpage

\end{document}